%% file: 00-main.tex
\documentclass[conference]{IEEEtran}
\IEEEoverridecommandlockouts
\usepackage{url}
\usepackage{multicol}
\usepackage{multirow}
\usepackage{amssymb}
\usepackage{pifont}
\usepackage{amsmath,amssymb,amsfonts}
\usepackage{amsmath,amssymb,amsfonts,color,soul}
\usepackage[ruled,vlined,linesnumbered]{algorithm2e}
\usepackage{algpseudocode}
\usepackage[flushleft]{threeparttable}
\usepackage{graphicx}
\usepackage{textcomp}
\usepackage{xcolor}

\def\BibTeX{{\rm B\kern-.05em{\sc i\kern-.025em b}\kern-.08em
    T\kern-.1667em\lower.7ex\hbox{E}\kern-.125emX}}

\usepackage[nolist]{acronym}
\acused{wifi}
\acused{wlan}
\acused{gnss}

\usepackage[backend=biber,style=ieee,  sortcites=true,isbn=false, doi=false, url=false,mincitenames=1, maxcitenames=1,maxbibnames=3,hyperref=false]{biblatex}
\usepackage{balance}

\usepackage{soul}
\soulregister\cite7
\soulregister\ref7
\soulregister\SI7
\soulregister\ac7

\usepackage[detect-all=true, binary-units=true]{siunitx}
\DeclareSIUnit\decibelm{dBm}
    
\usepackage[detect-all=true]{siunitx}
\usepackage{array}
\usepackage{booktabs}
\usepackage{mathtools}

\DeclarePairedDelimiter\abs{\lvert}{\rvert}%

\newcommand{\MYheader}{2022 International Conference on Localization and GNSS (ICL-GNSS), 7--9 June 2022, Tampere, Finland\\}

\newcommand{\MYcp}{978-1-6654-0575-1/22/\$31.00~\copyright~2022~IEEE}

\renewcommand{\MYheader}{}
\renewcommand{\MYcp}{}

\makeatletter
\def\ps@headings{%
	\def\@oddhead{}
	\def\@evenhead{}
	\def\@oddfoot{}%
	\def\@evenfoot{}}
\def\ps@IEEEtitlepagestyle{%
	\def\@oddhead{\hfill\MYheader\hfill}
	\def\@evenhead{\hfill\MYheader\hfill}
	\def\@oddfoot{\MYcp\hfill }%
	\def\@evenfoot{\MYcp\hfill}
    }
\makeatother
\begin{document}

\title{Lightweight Hybrid CNN-ELM Model for Multi-building and Multi-floor Classification

\thanks{%
Corresponding Author: D. Quezada Gaibor (\texttt{quezada@uji.com})
}
\thanks{The authors gratefully acknowledge funding from European Union’s Horizon 2020 Research and Innovation programme under the Marie Sk\l{}odowska Curie grant agreements No.~$813278$ (A-WEAR: A network for dynamic wearable applications with privacy constraints, {http://www.a-wear.eu/}) and No.~$101023072$ (ORIENTATE: Low-cost Reliable Indoor Positioning in Smart Factories, {http://orientate.dsi.uminho.pt}).}
}

\author{%
\IEEEauthorblockN{%
Darwin Quezada-Gaibor%
\IEEEauthorrefmark{1}\textsuperscript{,}\IEEEauthorrefmark{2}, %
Joaquín Torres-Sospedra%
\IEEEauthorrefmark{3}, %
\\Jari Nurmi%
\IEEEauthorrefmark{2}, %
Yevgeni Koucheryavy%
\IEEEauthorrefmark{2}, %
and Joaquín Huerta%
\IEEEauthorrefmark{1}%
}

\IEEEauthorblockA{\IEEEauthorrefmark{1}\textit{Institute of New Imaging Technologies}, \textit{Universitat Jaume I}, Castellón, Spain}
\IEEEauthorblockA{\IEEEauthorrefmark{2}\textit{Electrical Engineering Unit}, \textit{Tampere University}, Tampere, Finland}
\IEEEauthorblockA{\IEEEauthorrefmark{3}\textit{ALGORITMI Research Centre}, \textit{Universidade do Minho}, Guimarães, Portugal}
}

\maketitle

\begin{abstract}
Machine learning models have become an essential tool in current indoor positioning solutions, given their high capabilities to extract meaningful information from the environment. Convolutional neural networks (CNNs) are one of the most used neural networks (NNs) due to that they are capable of learning complex patterns from the input data. Another model used in indoor positioning solutions is the Extreme Learning Machine (ELM), which provides an acceptable generalization performance as well as a fast speed of learning. In this paper, we offer a lightweight combination of CNN and ELM, which provides a quick and accurate classification of building and floor, suitable for power and resource-constrained devices. As a result, the proposed model is $58\%$ faster than the benchmark, with a slight improvement in the classification accuracy (by less than $1\%$).
\end{abstract}


\input{acronyms}

\input{01-introduction}

\input{02-relatedwork}

\input{03-proposed-method}
\input{04-experiments-results}
\input{05-discussion}
\input{06-conclusions}


\balance

\renewcommand*{\UrlFont}{\rmfamily}\printbibliography

\end{document}

%% file: acronyms.tex
\begin{acronym}[XXX] 
\acro{ap}[AP]{Access Point}
\acro{apc}[APC]{affinity propagation clustering}
\acro{ble}[BLE]{Bluetooth Low Energy}
\acro{cnn}[CNN]{Convolutional Neural Network}
\acro{csi}[CSI]{Channel State Information}
\acro{dbscan}[DBSCAN]{Density-based Spatial Clustering of Applications with Noise}
\acro{elm}[ELM]{Extreme Learning Machine}
\acro{fp}[FP]{fingerprinting}
\acro{fpc}[FPC]{fingerprinting clustering}
\acro{gan}[GAN]{Generative Adversarial Network}
\acro{iot}[IoT]{Internet of Things}
\acro{ips}[IPS]{Indoor Positioning System}
\acro{knn}[$k$-NN]{k-nearest neighbors}
\acro{leakyrelu}[LeakyReLU]{Leaky Rectified Linear Unit}
\acro{lda}[LDA]{Linear Discriminant Analysis}
\acro{lbs}[LBS]{location-based service}
\acro{lstm}[LSTM]{Long short-term memory}
\acro{mac}[MAC]{Media Access Control}
\acro{ml}[ML]{Machine Learning}
\acro{mlp}[MLP]{multilayer perceptron}
\acro{nn}[NN]{Nearest Neighbour}
\acro{pca}[PCA]{Principal Component Analysis}
\acro{relu}[ReLU]{rectified linear}
\acro{rf}[RF]{Radio Frequency}
\acro{rnn}[RNN]{recurrent neural networks}
\acro{rp}[RP]{Reference Point}
\acro{rs}[RS]{Recommender Systems}
\acro{rss}[RSS]{Received Signal Strength}
\acro{sae}[SAE]{Stacked Auto-Encoder}
\acro{slfn}[SLFN]{Single hidden layer feedforward neural network}
\acro{svm}[SMV]{support vector machine}
\acro{tsne}[t-SNE]{T-distributed Stochastic Neighbor Embedding}
\acro{uwb}[UWB]{ultra-wideband}
\acro{vlc}[VLC]{Visible light communication}
\acro{wifi}[\mbox{Wi-Fi}]{IEEE 802.11 Wireless LAN}
\acro{wap}[WAP]{wireless access point}
\acro{wknn}[WKNN]{weighted k-nearest
neighbor}
\acro{wlan}[WLAN]{Wireless LAN}
\acro{wsn}[WSN]{Wireless Sensors Networks}

\end{acronym}

%% file: 01-introduction.tex
\section{Introduction}
\label{sec:introduction}

In the last two decades, the use of \ac{ml} algorithms in indoor positioning solutions are becoming more and more frequent, because of their high performance and accuracy. Thus, industry and academia are developing new \ac{ml} models to provide highly accurate solutions to the end-users. Some of these \ac{ml} models are already used in \ac{iot} and wearable devices \cite{daloia2020iot}. In this case, it is essential to keep a low computational load and high accuracy. There is, however, a trade-off between accuracy and computational complexity. The more accurate a model, the more computational resources used; finding an equilibrium between accuracy and power consumption has become a hot topic in \ac{ml}~\cite{brownlee2021exploring}.

Given the rapid growth of wearable and \ac{iot} devices that use positioning and localization services, it is essential to provide models that empower indoor positioning solutions in power-constrained devices. For instance, in \cite{torres2020new}, the authors provided three new variants for k-means clustering, which allowed a better distribution of \ac{wifi} fingerprints among the clusters, reducing the computational load (by approx. $40\%$) in comparison with the original K-means clustering. Other \ac{ml} algorithms have been used to enhance the positioning accuracy and/or floor hit rate, such as \ac{knn}~\cite{xue2018improved}, \ac{cnn}~\cite{shao2018indoor}, \ac{rnn}~\cite{hoang2019recurrent}, among others. However, these \ac{ml} algorithms may require high computational resources to be trained, being unsuitable to be deployed on power-constrained devices.

In order to reduce the training time, \cite{huang2006extreme} proposed a new learning algorithm for \ac{slfn} called \ac{elm}, which uses Moore-Penrose generalized inverse to compute the output weights of the neural network. Since this neural network does not use the traditional gradient-based learning algorithms, its training time is remarkably low. \ac{elm} has been widely used for classification, regression, clustering and dimensionality reduction, providing good general performance~\cite{yu2019active,li2018extreme,chen2020unsupervised,kasun2016dimension, lu2015robust}. Likewise, \cite{wu2016fast} proposed a new method based on \ac{svm} for classification and data undersampling to deal with unbalanced radio-maps. As a result, the author reduced the training and prediction time in the online phase by more than five times in comparison with the original algorithm (\ac{svm}).

This research combines a deep learning algorithm \ac{cnn} for feature learning and \ac{elm} to speed up the training and prediction stage. The aim is to provide an accurate and lightweight algorithm that can be used in power-constrained devices. Additionally, this combination allows learning the complex patterns of \ac{wifi} fingerprinting datasets in a more efficient way, improving the building and floor hit rate. 

The main contributions of this work are as follows:

\begin{itemize}
    \item An efficient combination of convolutional neural network (CNN) and extreme learning machine (ELM) for multi-building and multi-floor classification;
    \item Analyzing the proposed combination's classification performance for twelve \ac{wifi} fingerprinting datasets;
    \item Open-source code available for public usage~\cite{quezada2022supplementary}.
\end{itemize}

This research work is organized as follows. Section~\ref{sec:related_work} provides an overview of current studies in the field of interest. Section~\ref{sec:model} describes the model used in this article. Section~\ref{sec:usecases} shows the experiments carried out in this paper and their main results. Section~\ref{sec:discussion} provides a brief discussion of pros and cons of the proposed machine learning model. Finally, Section~\ref{sec:conclusion} provides the main conclusion of this work.

%% file: 02-relatedwork.tex
\section{Related work}
\label{sec:related_work}

\ac{wifi} fingerprinting is one of the most common techniques used for indoor positioning in commercial and open-source applications such as anyplace~\cite{mpeis2020anyplace}, FIND~\cite{website:find} and indoors~\cite{website:indoors}. Additionally, this technique has been widely studied during the last decade in order to provide accurate indoor positioning solutions. However, given the complexity of indoor environments, the accuracy of these applications may vary from one scenario to another. 

In order to extract meaningful information from \ac{wifi} fingerprinting datasets, some scholars have used \ac{ml} to learn complex patterns therefore reducing the positioning error. Additionally, to reduce the positioning error, it is essential to accurately estimate the building and floor (in the case of multi-building and multi-floor environments). However, if the complexity of the machine learning model increases, the computational load also increases during the training phase.

\ac{cnn} has been successfully used in multiple datasets for pattern recognition, being widely used for image segmentation and classification. Given its high performance, it has also been used in indoor positioning solutions. For instance, \citeauthor{song2019cnnloc} combined \ac{sae} neural network and \ac{cnn}, namely CNNLoc, for a better classification of fingerprints and accurately determine the building and floor. This combination was tested in two datasets, UJIIndoorLoc~\cite{torres2014ujiindoorloc} and Tampere dataset~\cite{lohan2017wifi}, obtaining $95.92\%$ and $94.13\%$ respectively, in the floor hit rate. Moreover, the authors got an accuracy of $100\%$ in the building hit rate in both datasets.

\ac{elm}-based algorithms have also been used to improve the accuracy of \acp{ips}. That is why \citeauthor{zou2015fast} proposed the use of the online sequential extreme learning machine (OS-ELM) algorithm to learn the environmental dynamics and, therefore, reduce the positioning error in comparison with the traditional \ac{elm}. Additionally, \ac{elm} network is used in indoor positioning applications given its fast training, and it was combined with other \ac{ml} algorithms.

\citeauthor{lian2019improved} combined \ac{elm} with \ac{knn} and adaptive weighted sparse representation classification (WSRC) namely A Fast-Accurate-Reliable Localization System (AFARLS). In the same fashion as \cite{song2019cnnloc}, AFARLS was tested in two public datasets: UJIIndoorLoc and Tampere datasets. As a result, the authors provided a robust algorithm resilient to changes in the size of datasets, outperforming different algorithms such as weighted centroid, \ac{rss} clustering and Log-Gaussian probability by more than $5\%$ in the floor hit rate and $6\%$ in the positioning accuracy.   

As can be observed, both \ac{elm} and \ac{cnn} have been widely tested in indoor applications offering better performance and fast training. However, to the authors' knowledge, they have never been explored together for indoor positioning. Accordingly, we propose to combine these two \ac{ml} algorithms in order to speed up the training and prediction stage as well as the fingerprints classification into floor and building.

%% file: 03-proposed-method.tex
\section{CNN-ELM Model}
\label{sec:model}

In this section, we provide a general description of the proposed \ac{cnn}-\ac{elm}.

\subsection{Data Preparation}

Data preparation is one of the fundamental steps prior to applying machine learning algorithms. Thus, data transformation, data scaling and data augmentation, among others, will directly impact the learning capabilities of a neural network. 

For \ac{wifi} fingerprinting, two different processes have been applied to the tested radio maps in order to reduce their complexity and improve their quality. The first step is to change the original format of the dataset by using \textit{powed} data representation, similar to \cite{song2019cnnloc, torres2022comprehensive}, as shown in Eq.~\ref{eq:powed}.

\begin{equation}
    \label{eq:powed}
    \begin{aligned}
     Powed_j(\textbf{X}) & =\begin{cases} 
        0, & \text{if $\ac{rss}_{j}=0$}, \\
        \left(\frac{\ac{rss}_j - min(\textbf{X})}{-min(\textbf{X})} \right)^e, & \text{otherwise}
    \end{cases}
    \end{aligned}
\end{equation}

where, $\textbf{X} \in \mathcal{R}^{N \times n}$ ($N$ number of fingerprints and $n$ number of \acp{ap} ) and represents a radio-map. $\ac{rss}_j$ is the signal strength indicator received from the j-$th$ \ac{ap} ($j=1,2,...,n$), $min(\textbf{X})$ is the lowest \ac{rss} value in the dataset and $e$ is the mathematical constant $e$.

The second step is data normalization (feature scaling). Here \textit{unit norm} normalization is applied to the data using Eq.~\ref{eq:unitnorm}.

\begin{equation}
    \label{eq:unitnorm}
    \widehat{\textbf{X}}_j = \frac{\textbf{X}_j}{\left\Vert \textbf{X}_j \right\Vert}, j=1,2,...,n
\end{equation}
where, $\widehat{\textbf{X}}_j$ is the normalized feature (\ac{ap}), $\textbf{X}_j$ represent the j-$th$ feature in the radio-map and $\left\Vert \textbf{X}_j \right\Vert$ is the Euclidean norm of $\textbf{X}_j$. 

\subsection{Convolutional Neural Network (CNN)}

As previously mentioned, indoor environments are considered one of the most challenging for positioning purposes. In the particular case of radio frequency-based indoor positioning technologies, the signals are affected by multiple factors, specially the multipath effect and signal fluctuation. These adverse factors are reflected in the \ac{rss} values and, therefore, in the position estimation increasing the positioning error. That is why some authors have proposed the use of deep neural networks such as \ac{cnn} in order to learn these fluctuations~\cite{song2019cnnloc,ibrahim2018cnn}.

The proposed feature learning model is composed of the following layers: a \ac{cnn} layer (\textit{Conv1D}) and a \textit{Pooling1D} layer. The \ac{cnn} layer is used to extract spatial characteristics of the radio map, and the Pooling1D reduces the dimensionality of the input by taking the average value over a spatial window of a pre-defined size (pool\_size). The $batch\_flatten$ layer converts each samples of the batch (4D and 3D) into a 2D data. All parameters used in the feature learning layers are listed in Table~\ref{table:cnn}.

\input{tables/tbl_cnn_parameters}

\subsection{Extreme Learning Machine (ELM)}
\ac{elm} is the learning method for \ac{slfn} networks, where the input weights and bias term are randomly generated and the output weights are analytically determined using \textit{Moore-Penrose Pseudoinverse}~\cite{huang2006extreme}. Considering $N$ arbitrary samples ($\textbf{x}_i, \textbf{t}_i$), where $\textbf{x}_i=[x_{i1}, x_{i2}, x_{i3},...,x_{in}]^T \in \mathcal{R}^n$ and $\textbf{t}_i=[t_{i1}, t_{i2}, t_{i3},...,t_{im}]^T \in \mathcal{R}^m$ ($i=1,2,...,N$). In the case of regression models, the objective is to find the relationship between the input ($\textbf{x}_i$) and the target ($\textbf{t}_i$). Given that both, input weights and bias term do not need to be tuned, the \ac{elm} is comparable to solving a least squares problem \cite{huang2006extreme, lu2016robust}. Thus, the first step is to map the inputs with the \ac{elm}'s hidden neurons onto a random feature space.

\begin{equation}
    \label{eq:mapfs}
    \textbf{h} : \textbf{x}_i \rightarrow \textbf{h}(\textbf{x}_i)
\end{equation}
where, $\textbf{h}(\textbf{x}_i) = h(\textbf{w}_{i} \cdot \textbf{x}_{j} + b_{i}) \in \mathcal{R}^{1 \times L}$. $\textbf{w}_i = [w_{i1}, w_{i2},...,w_{in}]^T \in \mathcal{R}^n$ are the initial input weights, which connect the input with the i-$th$ hidden neurons. $b_{i} \in \mathcal{R}$ is the bias term. $L$ is the number of hidden neurons. $\textbf{w}_{i} \cdot \textbf{x}_{j}$ is the dot product between $\textbf{w}_{i}$ and $\textbf{x}_{j}$~\cite{huang2006extreme}. We thus represent the hidden output layer (\textbf{H}) as follows:

\begin{gather}
\label{matrix:h}
\textbf{H}= 
\begin{bmatrix}
h(\textbf{w}_{1} \cdot \textbf{x}_{1} + b_{1}) & \hdots & h(\textbf{w}_{L} \cdot \textbf{x}_{1} + b_{L})\\
\vdots & \ddots & \vdots\\
h(\textbf{w}_{1} \cdot \textbf{x}_{N} + b_{1}) & \hdots & h(\textbf{w}_{L} \cdot \textbf{x}_{N} + b_{L})
\end{bmatrix}_{N \times L} 
\end{gather}

Thus, the output of the \ac{elm} is give by:

\begin{equation}
    \label{eq:elm}
    \textbf{T} = \textbf{H} \beta
\end{equation}
where, $\beta \in \mathcal{R}^{L \times m}$ represents the output weights of the \ac{elm}, which connect the hidden neurons with the output and $\textbf{T} \in \mathcal{R}^{N \times m}$ is the target matrix,

\begin{alignat}{3}
\beta=
\begin{bmatrix} 
\beta_1^T \\ \vdots \\ \beta_L^T \end{bmatrix} 
&& \text{ and } && 
\textbf{T}=
\begin{bmatrix} 
\textbf{t}_1^T \\ \vdots \\ \textbf{t}_N^T 
\end{bmatrix}
\end{alignat}

According to \cite{huang2006extreme}, the smallest norm least-squares solution of Eq.~\ref{eq:elmmp} can be achieved using \textit{Moore-Penrose Pseudoinverse} as follows:
\begin{equation}
    \label{eq:elmmp}
    \beta = \textbf{H}^{\dagger} \textbf{T}
\end{equation}
where, $\textbf{H}^{\dagger}$ is the \textit{Moore-Penrose Pseudoinverse} of $\textbf{H}$. When $H^T H$ is not singular $H^{\dagger}=(H^T H)^{-1} H^T$, otherwise, $H^{\dagger}=H^T(HH^T)^{-1}$. Additionally, a regularization term ($c$) has been added to the previous equation.

\begin{equation}
    \label{eq:elm2}
    \beta = \bigg(\textbf{H}^T \textbf{H}+\frac{1}{c}\bigg)^{-1} \textbf{H}^T \textbf{T}
\end{equation}

The \ac{elm} network works well with different activation functions such as sigmoid and sine, as was mentioned in~\cite{huang2006extreme}. In this research work, we use hyperbolic tangent sigmoid (tansig) as the main activation function.

\begin{equation}
    \label{eq:tansig}
    tansig = \frac{2}{(1+exp(-2*(\textbf{w}_{i} \cdot \textbf{x}_{j} + b_{i})))}-1 
\end{equation}

Finally, the 8-bit fixed-point quantization has been used in this implementation for power-constrained devices as in \cite{dogaru2019bconv}.

\begin{figure}[ht]
    \centering
        \includegraphics[width=0.5\textwidth]{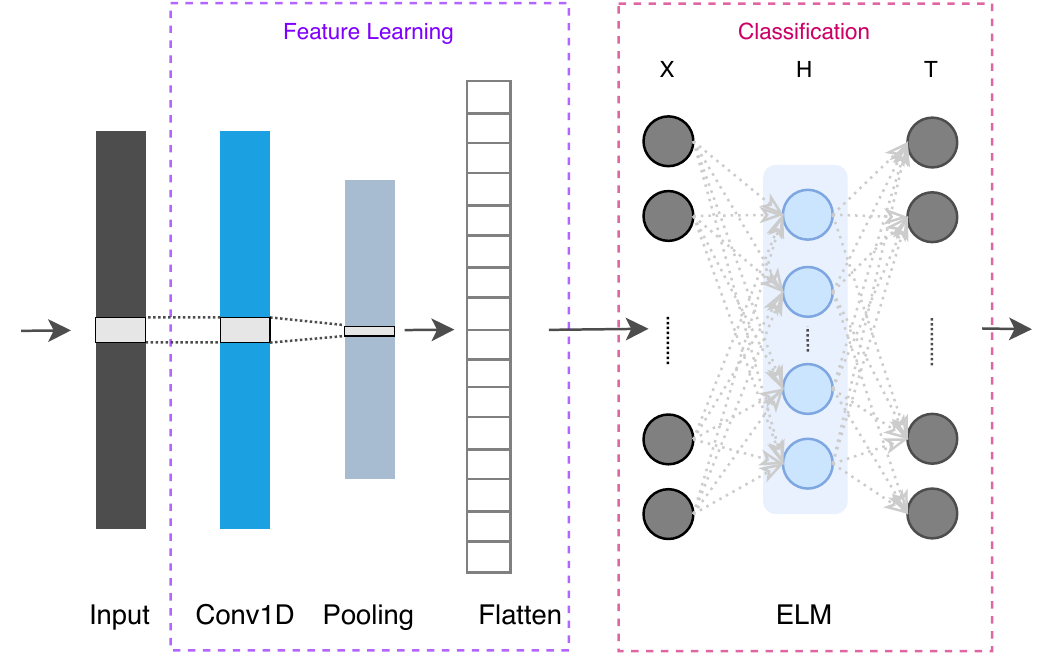}
    \caption{CNN-ELM model}
    \label{fig:cnnelm}
\end{figure}

Figure~\ref{fig:cnnelm} shows the combination of \ac{cnn} and \ac{elm} for fast building and floor classification. The first block represents the input data, then the feature learning block and finally the classification block (\ac{elm}). 

\subsection{\ac{cnn}-\ac{elm} Indoor Localisation}

\ac{wifi} fingerprinting is a popular indoor positioning technique given that \acp{ap} and/or \ac{wifi} routers are already deployed in both indoor and outdoor environments, avoiding deployment costs. Generally, this technique is divided into two phases. The off-line phase, where different \ac{rss} values are collected in known reference points in order to form the radio map. The formed radio map is divided into two or three sub-datasets to train \ac{ml} models to predict or classify the incoming fingerprints in the online phase. In the on-line phase, the user device obtains some \ac{rss} values (unknown position) which are used to predict the user location.

\begin{figure}[ht]
    \centering
        \includegraphics[width=0.5\textwidth]{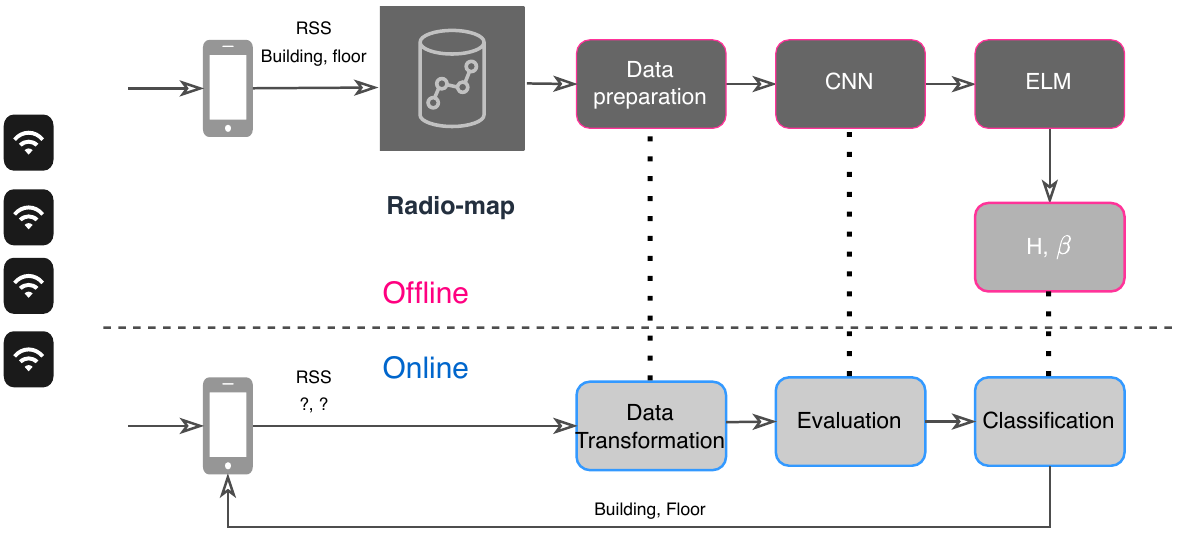}
    \caption{\ac{wifi} fingerprinting and CNN-ELM model}
    \label{fig:fp}
\end{figure}

Figure~\ref{fig:fp} shows the workflow at the off-line and on-line stages of the proposed \ac{cnn}-\ac{elm} model for smartphone-based \ac{wifi} fingerprinting. Since the proposed classification model for building and floor estimation does not require many computational resources in both off-line and on-line phase of \ac{wifi} fingerprinting, it can be used in power-constrained devices, servers with limited capabilities, and \ac{ips} with many concurrent users. 

%% file: tables/tbl_cnn_parameters.tex
\begin{table}[!hbtp]
    \tabcolsep 2.35pt
    
    \caption{Feature Learning Layers - Parameters}
    \label{table:cnn}
    \centering
    \begin{tabular}{l
    lc
    }
    \toprule
    Layer 
    &Parameter
    &Value\\
    \midrule

\multirow{4}{*}{Conv1D} & Padding & same\\
& Strides & 1 \\
& Filter & 2 \\
& Data format & channel\_last \\
\midrule
\multirow{5}{*}{Pooling1D} & strides & 2 \\
& Padding & valid \\
& pool\_mode & avg \\
& pool\_size & 2 \\
& Data format & channel\_last \\
\midrule
Activation funtion & & abs \\
\midrule
batch\_flatten & & \\
\bottomrule
    \end{tabular}
\end{table}

%% file: 04-experiments-results.tex
\section{Experiments and Results}
\label{sec:usecases}

This section provides a general description of the experiment setup, datasets used, and main results obtained in this research work.

\subsection{Experiment setup}

The experiments were performed using twelve public \ac{wifi} fingerprint datasets: UJIIndoorLoc (UJI~1--2)~\cite{torres2014ujiindoorloc}, LIB~1--2 \cite{MendozaSilva2018longterm} (Universitat Jaume I); 
TUT~1--7 \cite{ShresthaTalvitieEtAl13_Deconvolution,RazaviValkamaEtAl15_K,CramariucHuttunenEtAl16_Clustering,LohanTorres-SospedraEtAl17_Wi,RichterLohanEtAl18_WLAN,taudatasets} (Tampere University) and UTSIndoorLoc~\cite{song2019cnnloc} (University of Technology Sydney). These datasets are diverse and have been collected using multiple devices and in differing scenarios, such as libraries and universities. The proposed evaluation setup allows us to obtain a generalized assessment and meaningful results as presented by~\cite{9391692,torres2022comprehensive}. 

All the experiments have been carried out using a computer with the following characteristics: Intel® Core™ i7-8700T @ 2.40 GHz and 16 GB of RAM, the operating system is Fedora Linux 32, and the software used is python 3.9.

\input{tables/tbl_db_parameters}

Table~\ref{table:datasets} summarizes the characteristics of each dataset. $\abs{\mathcal{T}_{TR}}$ represents the number of samples in the training dataset, $\abs{\mathcal{T}_{TE}}$ is the number of samples in the test dataset, $\abs{\mathcal{A}}$ represents the number of \acp{ap}, DB Type shows if the dataset is multi-building (MB), multi-floor (MF) or both. $\abs{L}$ represent the number of hidden neurons used in the \ac{elm}.

\ac{knn} has been chosen as the baseline with $k=1$, and euclidean distance as the main distance metric. The \ac{knn} has been implemented using the KNeighborsClassifier class of sklearn library. Furthermore, CNNLoc~\cite{song2019cnnloc}, \ac{elm}, and AFARLS have been used to compare the performance of the proposed \ac{cnn}-\ac{elm} model in terms of building hit rate ($\zeta_{b}$), floor hit rate ($\zeta_{f}$), prediction time ($\delta_{tr}$) and training time ($\delta_{te}$). Thus, their normalized values $\tilde\zeta_{b}$, $\tilde\zeta_{f}$, $\tilde\delta_{tr}$ and $\tilde\delta_{te}$ are used to compare the results within the differing approaches. Given that \ac{knn} does not have any training stage, the CNNLoc training time was taken as baseline. 

In order to run the CNNLoc approach, the training dataset was divided into training and validation datasets.
First, the training dataset was divided into buildings (if dataset is multi-building) and then into floors. From each floor in each building, $10\%$ of fingerprints were randomly taken for validation. 

To choose the number of hidden nodes in the \ac{elm} network, the experiment was first run using five hidden neurons in the hidden layer and it was then increased in steps of five. The regularization term ($c$) in Eq.~\ref{eq:elm2} takes the values: $1$ (TUT~7), $0.1$ (TUT~1, TUT~6, UJI~1), $0.05$ (LIB~1, TUT~3, TUT~4) and $0.01$ (LIB~2, TUT~2, TUT~5, UJI~2, UTS~1). Finally, the number of neurons found to provide a good general performance were selected. Given the random components of \ac{cnn}-\ac{elm}, the random generation was seeded to ensure replicability.

The feature learning block was developed using \textit{Keras backend} for low-level tasks in order to reduce the load and computational time during training and testing.

\subsection{Results}

Table~\ref{table:comparison} shows the comparison between the results obtained by AFARLS~\cite{lian2019improved} and the proposed \ac{cnn}-\ac{elm} in UJI~1 and TUT~3, reveling that \ac{cnn}-\ac{elm} provides a slightly lower floor hit rate ($<4\%$) than AFARLS in both datasets. However, the number of hidden neurons used in our \ac{elm} is significantly lower --more than $47\%$-- than in AFARLS. Similarly, our proposed method provided significantly lower training and testing times, but $\delta_{tr}$ and $\delta_{te}$ reported in \cite{lian2019improved} include the 2D positioning error, which is not incorporated in this research. In the case of building hit rate, its accuracy was not affected.


\input{tables/tbl_uji_tamp}

\input{tables/tbl_results}

Table~\ref{table:results} shows the results obtained with \ac{knn}, CNNLoc, \ac{elm}, and the proposed \ac{cnn}-\ac{elm}. The baseline provides $100\%$ in the building hit rate in UJI~1 and UJI~2. The average accuracy in the floor hit rate is $95.24\%$, and the average classification time is \SI{0.52}{\second}. If we compare CNNLoc with the baseline, we see that the building hit rate slightly decreased in the UJI~1 dataset. Similarly, the average floor hit rate decreased by $2\%$ approximately. Nevertheless, the CNNLoc performance is better than the baseline in four datasets (LIB~1, TUT~5, TUT~6, and UJI~1). Surprisingly, the prediction time increased threefold compared with the baseline. 

The \ac{elm} network provides a better classification performance than CNNLoc, but it is still lower than the baseline. In the case of LIB~1, TUT~3 and TUT~7, the floor hit rates are slightly higher than the baseline (by less than $1\%$). There is a minimal reduction in the building hit rate in the UJI~1 and UJI~2 datasets. The training and testing time is, however, considerably lower in relation to the \ac{knn} and \ac{cnn}Loc in all datasets. For instance, the average testing time was reduced by more than $66\%$ in comparison with the baseline.

In the case of the \ac{cnn}-\ac{elm}, it has the same classification accuracy as the baseline ($1$ in the building hit rate and $1.0041$ in the floor hit rate). Compared to the CNNLoc, and \ac{elm} network, the classification accuracy of the \ac{cnn}-\ac{elm} increases by more than $2\%$ on average. In spite of the fact that there is a slight increment in the training and testing time, this is offset by improved classification accuracy when we compare our approach with \ac{elm} network. Thus, the average training time is more than sixty times faster than the CNNLoc, the average testing time was reduced by almost $58\%$ in contrast with baseline, and it is just $25\%$ higher than the \ac{elm} network.

\begin{figure}[ht]
    \centering
        \includegraphics[width=0.5\textwidth]{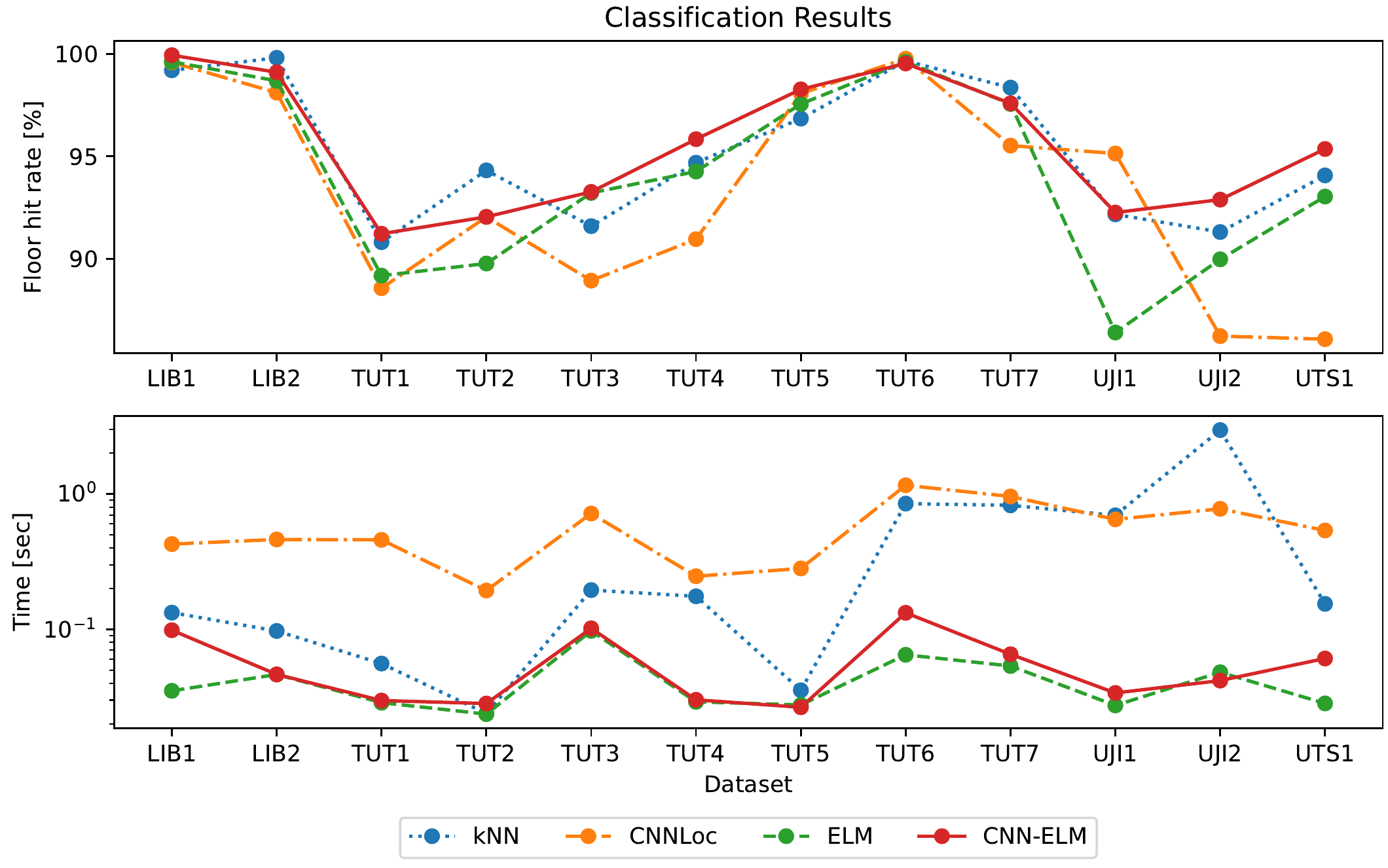}
    \caption{Classification results}
    \label{fig:classresults}
\end{figure}

Figure~\ref{fig:classresults} shows the classification results (without normalization) in terms of floor hit rate (top) and testing time (bottom). The minimum floor hit rate achieved with \ac{cnn}-\ac{elm} is greater than $91\%$ in the TUT~1 dataset, and the maximum is almost $100\%$ in LIB~1. Finally, \ac{cnn}-\ac{elm} provides better floor detection rate in UJI~2 and UTS~1, the largest datasets in terms of number of buildings and floors.

%% file: tables/tbl_db_parameters.tex
\begin{table}[!h]
    \tabcolsep 4pt
    \caption{Datasets Parameters}
    \label{table:datasets}
    \centering
    \begin{tabular}{l r r r c c c}
         \toprule
            Dataset 
            &$\abs{\mathcal{T}_{TR}}$
            &$\abs{\mathcal{T}_{TE}}$
            &$\abs{\mathcal{A}}$
            &$\abs{L}$
            & DB Type
            & Data Rep.
            \\

         \midrule
LIB~1 & 576 & 3120 & 174 & 105 & MF & Powed \\
LIB~2 & 576 & 3120 & 197 & 105 & MF & Powed \\
TUT~1 & 1476 & 490 & 309 & 75 & MF & Powed \\
TUT~2 & 584 & 176 & 354 & 160 & MF & Powed \\
TUT~3 & 697 & 3951 & 992 & 235 & MF & Powed \\
TUT~4 & 3951 & 697 & 992 & 275 & MF & Powed \\
TUT~5 & 446 & 982 & 489 & 195 & MF & Powed \\
TUT~6 & 3116 & 7269 & 652 & 450 & MF & Powed \\
TUT~7 & 2787 & 6504 & 801 & 200 & MF & Powed \\
UJI~1 & 19861 & 1111 & 520 & 530 & MB-MF & Powed \\
UJI~2 & 20972 & 5179 & 520 & 215 & MB-MF & Powed \\
UTS~1 & 9108 & 388 & 589 & 275 & MF & Powed \\
         \bottomrule
    \end{tabular}
    
\end{table}

%% file: tables/tbl_uji_tamp.tex
\begin{table}[!h]
    \tabcolsep 3pt
    \caption{Datasets Parameters}
    \label{table:comparison}
    \centering
    \begin{tabular}{l c c c c c c}
         \toprule
            \multirow{2}{*}{Approach} 
            &\multirow{2}{*}{Database}
            &\multirow{2}{*}{Parameters}
            &$\zeta_{b}$
            &$\zeta_{f}$
            &$\delta_{tr}$
            &$\delta_{te}$
            \\
            &&
            &$[\%]$
            &$[\%]$
            &$[\si{\sec}]$
            &$[\si{\sec}]$\\
         \midrule

\multirow{2}{*}{AFARLS~\cite{lian2019improved}}&UJI~1&$L=1000$&$100\%$&$95.41\%$&$84.68$&$0.21$\\
&TUT~3&$L=1000$&$-$&$94.18\%$&$2.40$&$0.57$\\
\midrule
\multirow{2}{*}{\ac{cnn}-\ac{elm}}&UJI~1&$L=530$&$100\%$&$92.26\%$&$0.26$&$0.03$\\
&TUT~3&$L=235$&$-$&$93.27\%$&$0.22$&$0.10$\\
\bottomrule
\end{tabular}
\end{table}

%% file: tables/tbl_results.tex
\begin{table*}[!hbtp]
    \tabcolsep 3.75pt
    
    \caption{Comparison among the 1-NN baseline, CNNLoc, ELM and CNN-ELM.}
    \label{table:results}
    \centering
    \resizebox{\textwidth}{!}{%
    \begin{tabular}{
    l
    cccccccc
    cccc
    cccc
    cccc
    }
         \toprule
            &\multicolumn{8}{c}{Baseline 1-NN}
            &\multicolumn{4}{c}{CNNLoc~\cite{song2019cnnloc}}
            &\multicolumn{4}{c}{ELM}
            &\multicolumn{4}{c}{CNN-ELM}
            \\
         \cmidrule(rl){2-9}
         \cmidrule(rl){10-13} 
         \cmidrule(rl){14-17}
         \cmidrule(rl){18-21}
            \multirow{2}{*}{Database}
            
            &\multicolumn{1}{c}{$\zeta_{b}$}
            &\multicolumn{1}{c}{$\zeta_{f}$}
            &\multicolumn{1}{c}{{$\delta_{tr}$}}
            &\multicolumn{1}{c}{$\delta_{te}$}
            &\multicolumn{1}{c}{$\tilde\zeta_{b}$}
            &\multicolumn{1}{c}{$\tilde\zeta_{f}$}
            &\multicolumn{1}{c}{$\tilde\delta{tr}$}
            &\multicolumn{1}{c}{$\tilde\delta_{te}$}

            &\multicolumn{1}{c}{$\tilde\zeta_{b}$}
            &\multicolumn{1}{c}{$\tilde\zeta_{f}$}
            &\multicolumn{1}{c}{$\tilde\delta{tr}$}
            &\multicolumn{1}{c}{$\tilde\delta_{te}$}
            
            &\multicolumn{1}{c}{$\tilde\zeta_{b}$}
            &\multicolumn{1}{c}{$\tilde\zeta_{f}$}
            &\multicolumn{1}{c}{$\tilde\delta{tr}$}
            &\multicolumn{1}{c}{$\tilde\delta_{te}$}

            &\multicolumn{1}{c}{$\tilde\zeta_{b}$}
            &\multicolumn{1}{c}{$\tilde\zeta_{f}$}
            &\multicolumn{1}{c}{$\tilde\delta{tr}$}
            &\multicolumn{1}{c}{$\tilde\delta_{te}$}\\

            &\multicolumn{1}{c}{$[\%]$}
            &\multicolumn{1}{c}{$[\%]$}
            &\multicolumn{1}{c}{$[\si{\sec}]$}
            &\multicolumn{1}{c}{$[\si{\sec}]$}
            &\multicolumn{1}{c}{$[-]$}
            &\multicolumn{1}{c}{$[-]$}
            &\multicolumn{1}{c}{$[-]$}
            &\multicolumn{1}{c}{$[-]$}
            
            &\multicolumn{1}{c}{$[-]$}
            &\multicolumn{1}{c}{$[-]$}
            &\multicolumn{1}{c}{$[-]$}
            &\multicolumn{1}{c}{$[-]$}
            
            &\multicolumn{1}{c}{$[-]$}
            &\multicolumn{1}{c}{$[-]$}
            &\multicolumn{1}{c}{$[-]$}
            &\multicolumn{1}{c}{$[-]$}

            &\multicolumn{1}{c}{$[-]$}
            &\multicolumn{1}{c}{$[-]$}
            &\multicolumn{1}{c}{$[-]$}
            &\multicolumn{1}{c}{$[-]$}\\
         \midrule

 LIB1 & - & 99.20 & - & 0.1328 & - & 1 & - & 1 & - & 1.0039 & 1 & 3.2084 & - & 1.0042 & 0.0105 & 0.2647 & - & 1.0074 & 0.0897 & 0.7417 \\ 
LIB2 & - & 99.81 & - & 0.0972 & - & 1 & - & 1 & - & 0.9830 & 1 & 4.7390 & - & 0.9888 & 0.0119 & 0.4769 & - & 0.9929 & 0.0244 & 0.4777 \\ 
TUT1 & - & 90.82 & - & 0.0559 & - & 1 & - & 1 & - & 0.9753 & 1 & 8.1930 & - & 0.9820 & 0.0042 & 0.5141 & - & 1.0045 & 0.0066 & 0.5328 \\ 
TUT2 & - & 94.32 & - & 0.0239 & - & 1 & - & 1 & - & 0.9759 & 1 & 8.0924 & - & 0.9518 & 0.0120 & 0.9895 & - & 0.9760 & 0.0147 & 1.1848 \\ 
TUT3 & - & 91.60 & - & 0.1949 & - & 1 & - & 1 & - & 0.9710 & 1 & 3.6773 & - & 1.0177 & 0.0138 & 0.4998 & - & 1.0182 & 0.0156 & 0.5218 \\ 
TUT4 & - & 94.69 & - & 0.1754 & - & 1 & - & 1 & - & 0.9606 & 1 & 1.4059 & - & 0.9954 & 0.0022 & 0.1661 & - & 1.0121 & 0.0042 & 0.1720 \\ 
TUT5 & - & 96.84 & - & 0.0355 & - & 1 & - & 1 & - & 1.0126 & 1 & 7.9418 & - & 1.0074 & 0.0121 & 0.7801 & - & 1.0147 & 0.0201 & 0.7493 \\ 
TUT6 & - & 99.66 & - & 0.8479 & - & 1 & - & 1 & - & 1.0011 & 1 & 1.3660 & - & 0.9996 & 0.0033 & 0.0765 & - & 0.9988 & 0.0079 & 0.1562 \\ 
TUT7 & - & 98.36 & - & 0.8233 & - & 1 & - & 1 & - & 0.9712 & 1 & 1.1628 & - & 0.9919 & 0.0029 & 0.0651 & - & 0.9922 & 0.0052 & 0.0795 \\ 
UJI1 & 100 & 92.17 & - & 0.6946 & 1 & 1 & - & 1 & 0.9973 & 1.0322 & 1 & 0.9338 & 0.9991 & 0.9375 & 0.0007 & 0.0395 & 1 & 1.0010 & 0.0010 & 0.0488 \\ 
UJI2 & 100 & 91.31 & - & 2.9602 & 1 & 1 & - & 1 & 1 & 0.9444 & 1 & 0.2622 & 0.9996 & 0.9854 & 0.0005 & 0.0163 & 1 & 1.0173 & 0.0011 & 0.0141 \\ 
UTS1 & - & 94.07 & - & 0.1541 & - & 1 & - & 1 & - & 0.9151 & 1 & 3.4835 & - & 0.9890 & 0.0011 & 0.1840 & - & 1.0137 & 0.0019 & 0.3950 \\ 
\midrule
Avg. & 100 & 95.24 & - & 0.52 & 1 & 1 & - & 1 & 0.9987 & 0.9789 & 1 & 3.7055 & 0.9994 & 0.9876 & 0.0063 & 0.3394 & 1 & 1.0041 & 0.0160 & 0.4228 \\

         \bottomrule
    \end{tabular}
    }
\end{table*}

%% file: 05-discussion.tex
\section{Discussion}
\label{sec:discussion}

In the previous section we compared four different approaches; \ac{knn}, CNNLoc, \ac{elm} and, briefly, AFARLS with our approach (\ac{cnn}-\ac{elm}) in Table~\ref{table:comparison}. Although, the classification accuracy with \ac{cnn}-\ac{elm} model is not as high as AFARLS in the case of the UJI~1 (UJIIndoorLoc in~\cite{lian2019improved}) and TUT~3 (Tampere in \cite{lian2019improved}) datasets. The training and testing time is significantly lower than AFARLS.

We have to consider that data preprocessing is fundamental prior to applying any machine learning model. This case is not the exception, \textit{powed} data representation and data normalization technique allowed to enhance the dataset's characteristics. These two techniques were essential to achieve better classification accuracy and processing time (training and testing time).

Similarly, code and algorithm optimization play an important role in offering fast training and testing time. Thus, if additional calculations are done during the prediction time, the time response will increase along with the computational load. An inefficient implementation may raise scalability problem when deploying an Indoor Positioning system. 

As expected, the feature learning block allowed us to extract a high level of characteristics from the radio map in such a manner that the \ac{elm} network can process that information more efficiently. Thus, with minimal training time, the network can provide high levels of accuracy in line with to complex networks such as CNNLoc or AFARLS.

%% file: 06-conclusions.tex
\section{Conclusions}
\label{sec:conclusion}

In this paper, we present a lightweight ensemble \ac{cnn}-\ac{elm} model for building and floor indoor-positioning classification. 

We have performed a comprehensive evaluation using twelve diverse datasets and compared against three known models from the literature (CNNLoc, \ac{knn} and \ac{elm}). Two of these datasets were also compared against AFARLS approach.

Although the \ac{cnn}-\ac{elm} network is simple, it has been capable of providing similar or better performance than complex machine learning models in short run times. Thus, the average testing time was $58\%$ faster than the baseline model based on $1$-NN, providing almost the same positioning error. Compared with the AFARLS approach, the classification accuracy of \ac{cnn}-\ac{elm} was less than $4\%$ worse. However, the number of hidden neurons used in the \ac{elm} was $\approx 47\%$ less than the number of hidden neurons used in the AFARLS architecture in the datasets analysed. This makes our proposal also lighter to operate in terms of memory requirements. 

Future work will include analysis of using the \ac{cnn}-\ac{elm} network to provide 2D and 3D positioning estimation as well as new optimisation techniques.